\documentclass{aa}

\usepackage{natbib}
\bibpunct{(}{)}{;}{a}{}{,}

\usepackage{txfonts}

\usepackage[dvips]{graphicx}
\usepackage{subfigure}

\begin{document}

\def\lef{{\it Left: }}
\def\rig{{\it Right: }}
\def\top{{\it Top: }}
\def\bot{{\it Bottom: }}
\def\mid{{\it Middle: }}

\authorrunning{R. Da Silva et al.}

\title{Elodie metallicity-biased search for transiting Hot Jupiters
\thanks{Based on radial velocities collected with the ELODIE spectrograph
mounted on the 193-cm telescope at the Observatoire de Haute Provence,
France. Additional observations were made using the new SOPHIE
spectrograph (run 06B.PNP.CONS) that replaces ELODIE.}}
\subtitle{IV. Intermediate period planets orbiting the stars HD\,43691 and
HD\,132406}

\author{R. Da Silva\inst{1}
\and S. Udry\inst{1}
\and F. Bouchy\inst{3}
\and C. Moutou\inst{2}
\and M. Mayor\inst{1}
\and J.-L. Beuzit\inst{5}
\and X. Bonfils\inst{5}
\and X. Delfosse\inst{5}
\and M. Desort\inst{5}
\and T. Forveille\inst{5}
\and F. Galland\inst{5}
\and G. H\'ebrard\inst{3}
\and A.-M. Lagrange\inst{5}
\and B. Loeillet\inst{2}
\and C. Lovis\inst{1}
\and F. Pepe\inst{1}
\and C. Perrier\inst{5}
\and F. Pont\inst{1}
\and D. Queloz\inst{1}
\and N.C. Santos\inst{4,1}
\and D. S\'egransan\inst{1}
\and J.-P. Sivan\inst{2}
\and A. Vidal-Madjar\inst{3}
\and S. Zucker\inst{6}}

\offprints{R. Da Silva,\\
\email{Ronaldo.daSilva@obs.unige.ch}}

\institute{
Observatoire Astronomique de l'Universit\'e de Gen\`eve, 1290 Sauverny,
Switzerland
\and
Laboratoire d'Astrophysique de Marseille, UMR6110 CNRS, Universit\'e de
Provence, Traverse du Siphon, BP8, 13376 Marseille Cedex 12, France
\and
Institut d'Astrophysique de Paris, UMR7095 CNRS, Universit\'e Pierre \&
Marie Curie, 98bis bd Arago, 75014 Paris, France
\and
Centro de Astrof\'\i sica, Universidade do Porto, Rua das Estrelas,
4150-762 Porto, Portugal
\and
Laboratoire d'Astrophysique de Grenoble, BP 53X, 38041 Grenoble Cedex,
France
\and
Department of Geophysics and Planetary Sciences, Raymond and Beverly
Sackler Faculty of Exact Sciences, Tel Aviv University, Tel Aviv 69978,
Israel
}

\date{Received / accepted}
%
%
\abstract{
We report here the discovery of two planet candidates as a result of our
planet-search programme biased in favour of high-metallicity stars, using
the ELODIE spectrograph at the Observatoire de Haute Provence. One of them
has a minimum mass $m_2\sin{i}$ = 2.5~M$_{\rm Jup}$ and is orbiting the
metal-rich star \object{HD\,43691} with period $P$ = 40~days and
eccentricity $e$~=~0.14. The other planet has a minimum mass $m_2\sin{i}$ =
5.6~M$_{\rm Jup}$ and orbits the slightly metal-rich star
\object{HD\,132406} with period $P$~=~974~days and eccentricity $e$~=~0.34.
Both stars were followed up with additional observations using the new
SOPHIE spectrograph that replaces the ELODIE instrument, allowing an
improved orbital solution for the systems.
\\
}

\maketitle
%
%
\section{Introduction}
After the first publications suggesting the metal-rich nature of stars
hosting giant planets in close orbits \citep{Gonzalez1997,Gonzalez1998},
a number of works concerning the theme came out in the last few years
\citep{Santosetal2001,Santosetal2004,FischerValenti2005,Gonzalez2006}. With
the increasing number of known planets, statistical studies were able to
verify that the frequency of stars hosting a planetary companion is highly
correlated with the stellar metallicity. The results show that the
probability of finding a close-in giant planet is about 25-30\% for the most
metal-rich stars ([Fe/H] $>$ 0.3) and only 3\% for stars with solar
metallicity. The extrasolar planet search biased in favour of
high-metallicity stars can thus more quickly conduct to the discovery of
planets in short-period orbits, the so-called hot Jupiters ($P < 10$~days),
increasing the chances of finding planetary transits. The identification of
planets transiting bright stars provides a powerful approach to determine
fundamental constraints on the mechanisms of planet formation, the physical
properties of the exoplanet, and the geometry of the system.

Based on these assumptions, a few programmes have been initiated with the
approach of looking for planets orbiting high-metallicity stars. One of them
is the N2K consortium \citep{Fischeretal2004}, which monitors nearly 2000
main-sequence and subgiant stars. Another project was conducted by our team
with the ELODIE spectrograph at the Observatoire de Haute Provence
\citep{DaSilvaetal2006}. From a sample of more than a thousand solar-type
stars, we selected the more metallic ones after the first measurement to
monitor their radial velocities.

Our programme has already yielded the detection of four hot Jupiters, with
periods between 2.2 and 6.8~days and minimum masses between 1.0 and 2.1
${\rm M_{Jup}}$, orbiting the stars \object{HD\,118203} and
\object{HD\,149143}\footnote{The planet around HD\,149143 was independently
discovered by \citet{Fischeretal2006} and the one orbiting HD\,185269 was
also published by \citet{Johnsonetal2006}.} \citep{DaSilvaetal2006},
\object{HD\,189733} \citep{Bouchyetal2005}, and \object{HD\,185269}$^1$
\citep{Moutouetal2006}.

For the star HD\,189733, additional photometric measurements have led to the
observation of a planetary transit, making possible the determination of
some parameters of the companion, such as mass, radius and mean density
\citep{Bouchyetal2005,Bakosetal2006,Winnetal2007}. This system is our best
result, a very good example of what we expect to obtain with our programme,
and revealed to be of particular interest for further studies.
\citet{Demingetal2006} analysed the infrared thermal emission during an
eclipse of HD\,189733\,b using the {\it Spitzer Space Telescope}
\citep{Werneretal2004} and determined the brightness temperature of the
planet at 16~$\mu$m. \citet{Knutsonetal2007}, doing observations with
{\it Spitzer} at $8~\mu$m, were able to construct a map of the temperature
distribution of HD\,189733\,b, estimating a minimum and a maximum brightness
temperature at this wavelength. \citet{FortneyMarley2007} analysed the mid
infrared observations of HD\,189733 and suggested a possible presence of
water vapor in the atmosphere of the planetary companion. Observations with
the {\it Hubble Space Telescope} have also been proposed in order to
perform precise measurements of the size and the orbital inclination angle
of HD\,189733\,b (Pont et al., in preparation).

In this paper, we report the discovery of two new planet candidates,
resulted from our ELODIE planet search programme biased towards metal-rich
stars: a 2.5~Jupiter-mass planet orbiting the star HD\,43691 with period
$P$~=~40~days, and a 5.6~Jupiter-mass planet in a long-period orbit of
$P$~=~974~days around the star HD\,132406. Such results are complemented by
additional measurements made using SOPHIE \citep{Bouchyetal2006}, the new
spectrograph that replaces ELODIE.

The radial velocity observations that have led to these results are
described in Sect.~\ref{obs}. The observed and derived parameters of the
star HD\,43691 together with the orbital solution adopted are presented in
Sect.~\ref{star_par1}. The same are presented in Sect.~\ref{star_par2} for
the star HD\,132406. In Sect.~\ref{conc} we discuss the present and future
status of the observational programme.
%
%
\section{Observations}
\label{obs}
The stars HD\,43691 and HD\,132406 are both targets in our "ELODIE
metallicity-biased search for transiting hot Jupiters" survey
\citep{DaSilvaetal2006}, conducted from March 2004 until August 2006 with
the ELODIE spectrograph \citep{Baranneetal1996} on the 193-cm telescope at
the Observatoire de Haute Provence (France). In this programme we
essentially searched for Jupiter-like planets orbiting metal-rich stars,
assuming that such stars are more likely to host giant planets.

After obtaining the first spectrum of HD\,43691 and HD\,132406, we have
verified the high metallicity of these stars from a calibration of the
surface of the ELODIE cross-correlation functions \citep{Santosetal2002,
Naef2003}. After three measurements, we could clearly see in both stars a
significant radial velocity variation. We therefore conducted follow-up
observations with ELODIE, and we obtained 22 spectra of HD\,43691 from
November 2004 (JD~=~2\,453\,333) to May 2006 (JD~=~2\,453\,872), and 17
spectra of HD\,132406 from May 2004 (JD~=~2\,453\,152) to June 2006
(JD~=~2\,453\,900).

The ELODIE instrument was decommissioned in August 2006 and replaced by the
SOPHIE spectrograph. Additional measurements were then obtained using this
new instrument: 14 spectra of HD\,43691 from November 2006
(JD~=~2\,454\,044) to February 2007 (JD~=~2\,454\,155), and 4 spectra of
HD\,132406 from December 2006 (JD~=~2\,454\,080) to May 2007
(JD~=~2\,454\,230).

\begin{table}[t!]
\centering
  \caption[]{ELODIE and SOPHIE radial velocities of HD\,43691. All values
             are relative to the solar system barycentre. The uncertainties
	     correspond to the photon-noise errors.}
  \label{rad_vel1}
\begin{tabular}{ccc}
\hline
\hline
\noalign{\smallskip}
JD $-$ 2\,400\,000 & RV              & Uncertainty     \\
$[$days$]$         & $[$km\,s$^{-1}]$ & $[$km\,s$^{-1}]$ \\
\noalign{\smallskip}
\hline
\multicolumn{3}{c}{ELODIE measurements} \\
\hline
\noalign{\smallskip}
53333.6255 & $-$29.123 & 0.011 \\
53337.6057 & $-$29.098 & 0.012 \\
53398.4118 & $-$29.015 & 0.012 \\
53690.6694 & $-$28.945 & 0.013 \\
53692.6430 & $-$28.962 & 0.013 \\
53693.6240 & $-$28.983 & 0.014 \\
53714.5653 & $-$28.912 & 0.014 \\
53715.5590 & $-$28.903 & 0.015 \\
53718.5549 & $-$28.904 & 0.013 \\
53719.5681 & $-$28.885 & 0.010 \\
53720.5323 & $-$28.853 & 0.017 \\
53721.5123 & $-$28.892 & 0.018 \\
53722.5237 & $-$28.907 & 0.015 \\
53728.4315 & $-$28.969 & 0.018 \\
53749.5333 & $-$28.989 & 0.014 \\
53750.4982 & $-$28.958 & 0.014 \\
53756.4444 & $-$28.840 & 0.018 \\
53808.2808 & $-$29.091 & 0.012 \\
53809.2850 & $-$29.087 & 0.009 \\
53839.3031 & $-$28.960 & 0.019 \\
53870.3405 & $-$28.881 & 0.017 \\
53872.3456 & $-$28.923 & 0.020 \\
\noalign{\smallskip}
\hline
\multicolumn{3}{c}{SOPHIE measurements} \\
\hline
\noalign{\smallskip}
54044.6274 & $-$28.980 & 0.003 \\
54051.6411 & $-$28.830 & 0.004 \\
54053.5968 & $-$28.854 & 0.003 \\
54078.6009 & $-$29.059 & 0.004 \\
54079.4859 & $-$29.040 & 0.004 \\
54080.4572 & $-$29.018 & 0.004 \\
54081.4440 & $-$28.993 & 0.003 \\
54087.4744 & $-$28.858 & 0.004 \\
54088.5858 & $-$28.872 & 0.004 \\
54089.6137 & $-$28.846 & 0.004 \\
54142.4798 & $-$29.064 & 0.004 \\
54148.4607 & $-$29.089 & 0.004 \\
54151.4010 & $-$29.065 & 0.004 \\
54155.4284 & $-$28.981 & 0.004 \\
\noalign{\smallskip}
\hline
\end{tabular}
\end{table}

With ELODIE, the average signal-to-noise ratio (S/N) calculated from the
spectra at $\lambda5500$~{\AA} is $\sim$40 for both stars, with a typical
exposure time of 20~min. On the other hand, the gain in efficiency of
SOPHIE compared to ELODIE is more than one magnitude in the High-Resolution
mode (used for high precision radial-velocity measurements). Typical S/N
obtained with SOPHIE are thus twice larger than those of ELODIE for
exposure times 2-3 times smaller. Table~\ref{rad_vel1} lists the radial
velocities of HD\,43691 and Table~\ref{rad_vel2} lists those of HD\,132406.

Following \citet{ZuckerMazeh2001}, we tried to look for the astrometric
signatures of the two orbits in Hipparcos Intermediate Astrometric Data
(IAD). HD132406, whose best-fit Keplerian period was close to the Hipparcos
mission duration, seemed especially suitable for this kind of analysis. We
found no evidence of astrometric signatures. Furthermore, the mass upper
limits that the IAD produce are in the stellar regime and therefore do not
provide any useful constraint.

\begin{table}[t!]
\centering
  \caption[]{ELODIE and SOPHIE radial velocities of HD\,132406. All values
             are relative to the solar system barycentre. The uncertainties
	     correspond to the photon-noise errors.}
  \label{rad_vel2}
\begin{tabular}{ccc}
\hline
\hline
\noalign{\smallskip}
JD $-$ 2\,400\,000 & RV              & Uncertainty     \\
$[$days$]$         & $[$km\,s$^{-1}]$ & $[$km\,s$^{-1}]$ \\
\noalign{\smallskip}
\hline
\multicolumn{3}{c}{ELODIE measurements} \\
\hline
\noalign{\smallskip}
53152.4773 & $-$37.821 & 0.010 \\
53154.4825 & $-$37.837 & 0.008 \\
53218.3605 & $-$37.858 & 0.010 \\
53520.4238 & $-$37.928 & 0.013 \\
53536.4140 & $-$37.875 & 0.010 \\
53576.3681 & $-$37.859 & 0.012 \\
53596.3805 & $-$37.818 & 0.014 \\
53807.6666 & $-$37.727 & 0.025 \\
53808.6537 & $-$37.755 & 0.011 \\
53809.6643 & $-$37.742 & 0.011 \\
53869.5117 & $-$37.779 & 0.011 \\
53870.4406 & $-$37.771 & 0.007 \\
53873.4386 & $-$37.768 & 0.009 \\
53895.4251 & $-$37.757 & 0.009 \\
53896.4387 & $-$37.745 & 0.012 \\
53899.4286 & $-$37.743 & 0.019 \\
53900.4376 & $-$37.770 & 0.013 \\
\noalign{\smallskip}
\hline
\multicolumn{3}{c}{SOPHIE measurements} \\
\hline
\noalign{\smallskip}
54080.7252 & $-$37.718 & 0.003 \\
54173.6848 & $-$37.752 & 0.004 \\
54187.6332 & $-$37.770 & 0.004 \\
54230.5803 & $-$37.790 & 0.004 \\
\noalign{\smallskip}
\hline
\end{tabular}
\end{table}

In order to derive some of the fundamental stellar parameters, like
effective temperature, surface gravity and metallicity, using accurate
spectroscopic analysis, we obtained a high S/N spectrum ($\sim$130 at
$\lambda5500$~{\AA}) of HD\,43691 with the SOPHIE spectrograph.

%
%
\section{A planetary companion to HD\,43691}
\label{star_par1}
\subsection{Stellar characteristics of HD\,43691}
\label{star_car1}

\begin{table}[t!]
\centering
  \caption[]{Observed and estimated parameters of HD\,43691 and HD\,132406.
             Some of the stellar parameters of HD\,43691 were obtained from
	     spectroscopic analysis while those of HD\,132406 come from
	     calibrations of the ELODIE CCF.}
  \label{stellar_par}
\begin{tabular}{lccl}
\hline
\hline
\noalign{\smallskip}
              & HD\,43691        & HD\,132406       &		     \\
\noalign{\smallskip}
\hline
\noalign{\smallskip}
Spectral Type & G0~IV            & G0~V		    &		     \\
$V$           & 8.03             & 8.45 	    &		     \\
$B-V$         & 0.596            & 0.65 	    &		     \\
$\pi$         & 10.73 $\pm$ 1.16 & 14.09 $\pm$ 0.77 & [mas]	     \\
$M_V$         & 3.18             & 4.19 	    &		     \\
BC            & $-0.034$         & $-0.062$	    &		     \\
$T_{\rm eff}$ & 6200 $\pm$ 40    & 5885 $\pm$ 50    & [K]	     \\
$M_\star$     & 1.38 $\pm$ 0.05  & 1.09 $\pm$ 0.05  & M$_{\odot}$    \\
{\it age}     & 2.0 - 3.6        & 6.4  $\pm$ 0.8   & Gyr	     \\
log~$g$       & 4.28 $\pm$ 0.13  &                  &                \\
$[$Fe/H$]$    & 0.28 $\pm$ 0.05  & 0.18 $\pm$ 0.05  &		     \\
$v{\sin i}$   & 4.7              & 1.7  	    & [km\,s$^{-1}$] \\
\noalign{\smallskip}
\hline
\end{tabular}
\end{table}

HD\,43691 (HIP\,30057) is listed in the Hipparcos catalogue \citep{ESA1997} 
as a G0 star in the northern hemisphere with visual magnitude V = 8.03,
color index $B-V$ = 0.596 and parallax $\pi$ = 10.73 $\pm$ 1.16 mas (a
distance of 93~pc from the Sun). The bolometric correction is BC = $-$0.034,
derived from \citet{Flower1996}. Using the Hipparcos parameters we derived
an absolute magnitude $M_V$ = 3.18, which represents a high luminosity for a
G0 star. This suggests that HD\,43691 is slightly evolved towards the
subgiant branch. \citet{Nordstrometal2004} found a difference of 1.19 mag
from the ZAMS, indicating the degree of evolution of this star.

Applying the spectroscopic analysis described in \citet{Santosetal2004} to
the high S/N spectrum of HD\,43691 we obtained: $T_{\rm eff}$ = 6200 $\pm$
40~K, log~$g$ = 4.28 $\pm$ 0.13 and [Fe/H]~=~0.28~$\pm$~0.05. From the
calibrations of the ELODIE CCF \citep{Santosetal2002,Naef2003}, we estimated
a slightly smaller but compatible value for the metallicity ([Fe/H] = 0.22
$\pm$ 0.05), and a projected rotation velocity $v{\sin i}$ =
4.7~km\,s$^{-1}$.

With these stellar parameters, we estimated the mass and age of HD\,43691
using the Geneva models of stellar evolution computed by
\citet{Schaereretal1993}. We found a mass of $M_\star$ = 1.38 $\pm$
0.05~M$_{\odot}$ and an age between 2.0 and 3.6~Gyr, which are in agreement
with the determinations done by \citet{Nordstrometal2004}: mass $M_\star$ =
1.38 $\pm$ 0.08~M$_{\odot}$ and age 2.6 $\pm$ 0.5~Gyr. These values are
compatible with the star being slightly evolved, especially taking into
account the stellar metallicity \citep{Mowlavietal1998}. The observed and
derived stellar parameters of HD\,43691 are shown in
Table~\ref{stellar_par}.

%
%
\subsection{Orbital solution for HD\,43691\,b}
\begin{figure}[t!]
\centering
  \begin{minipage}[t]{0.45\textwidth}
    \centering
    \resizebox{0.98\hsize}{!}{\includegraphics{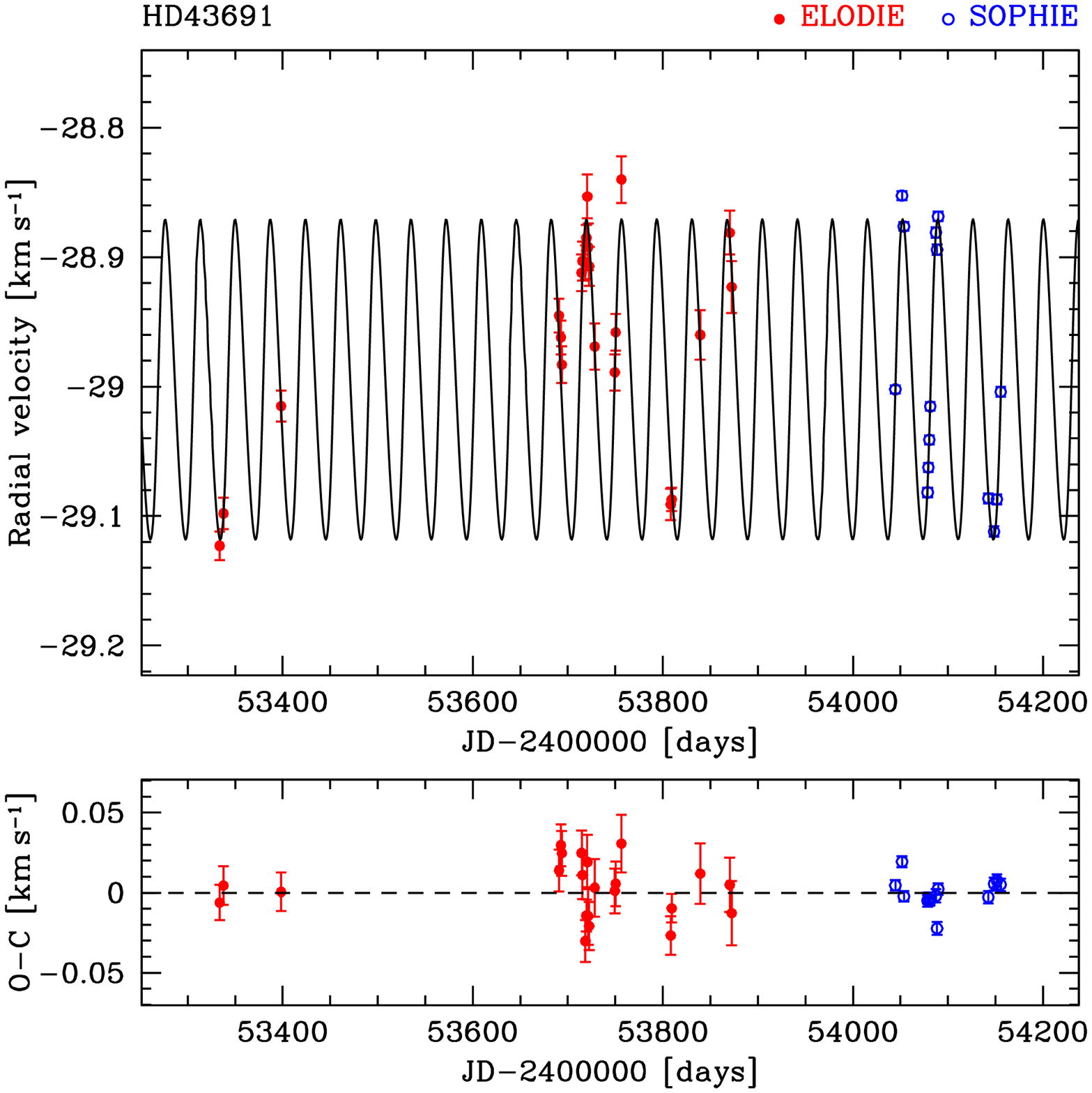}}
  \end{minipage} \\
  \begin{minipage}[t]{0.45\textwidth}
    \centering
    \resizebox{0.98\hsize}{!}{\includegraphics{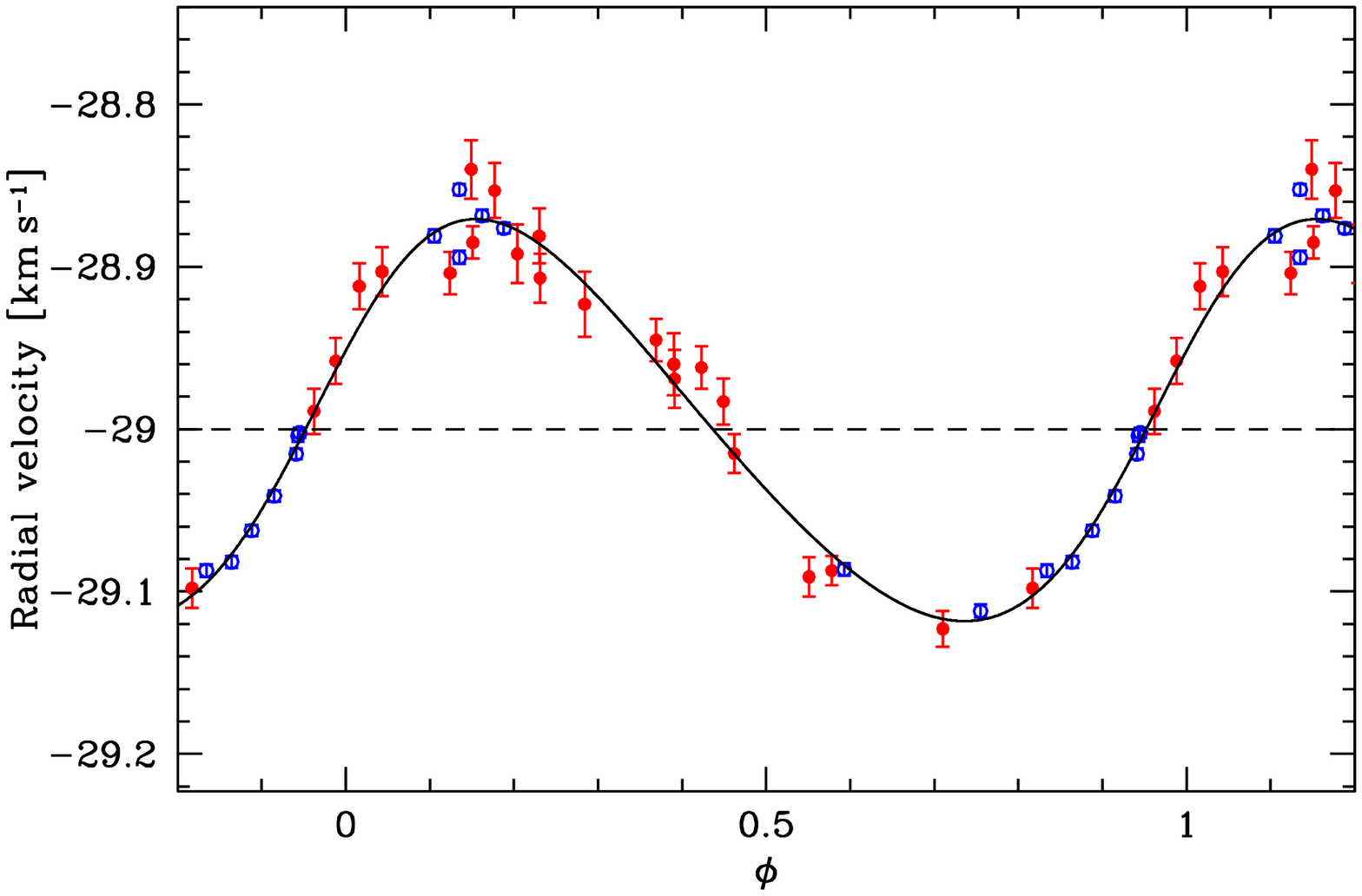}}
  \end{minipage}%
  \caption{\top ELODIE and SOPHIE radial velocities of HD\,43691 plotted
           together with the best Keplerian solution that fits the combined
	   measurements.
	   \mid Residuals around the solution, with
	   $\sigma_{\rm E}$~=~17.5~m\,s$^{-1}$ for ELODIE,
	   $\sigma_{\rm S}$~=~9.0~m\,s$^{-1}$ for SOPHIE, and
	   $\sigma_{\rm ES}$~=~10.0~m\,s$^{-1}$ for the combined data
	   points.
	   \bot Phase-folded radial velocities with the best Keplerian
	   solution. Error bars represent the photon-noise errors.}
  \label{rv_1}
\end{figure}

The best Keplerian orbital solution fitted to the radial velocities of
HD\,43691, using both ELODIE and SOPHIE observations, provides a orbit with
period $P$~=~36.96~$\pm$~0.02~days and eccentricity $e$ = 0.14 $\pm$ 0.02.
With the estimated value for the primary mass of 1.38~M$_{\odot}$ we
obtained a minimum mass $m_2 {\sin i}$ = 2.49~M$_{\rm Jup}$ and a
separation of 0.24~AU for the planetary companion. The solution includes a
velocity zero-point of the two datasets as a free parameter, and the
difference between them is $\Delta_{\rm S-E}~=~23~\pm~4$~m\,s$^{-1}$. For
this solution, we estimated a false alarm probability of \mbox{$1.3 \times
10^{-5}$} using the approach described in \citet{HorneBaliunas1986}.
 
\begin{table}[t]
\centering
  \caption[]{Orbital elements for the best Keplerian solution of HD\,43691
             and HD\,132406 as well as the inferred planetary parameters.}
  \label{orb_elem}
\begin{tabular}{l r@{ }l r@{ }l l}
\hline
\hline
\noalign{\smallskip}
 & \multicolumn{2}{c}{HD\,43691} & \multicolumn{2}{c}{HD\,132406} & \\
\noalign{\smallskip}
\hline
\noalign{\smallskip}
$P$                  & 36.96     & $\pm$ 0.02  & 974       & $\pm$ 39    & [days]                \\
$T$                  & 54046.6   & $\pm$ 0.5   & 53474     & $\pm$ 44    & {\scriptsize [JD $-$ 2\,400\,000]}  \\
$e$                  & 0.14      & $\pm$ 0.02  & 0.34      & $\pm$ 0.09  &                       \\
$V$                  & $-$29.000 & $\pm$ 0.003 & $-$37.840 & $\pm$ 0.008 & [km\,s$^{-1}$]        \\
$\omega$             & 290       & $\pm$ 5     & 214       & $\pm$ 19    & [deg]                 \\
$K$                  & 124       & $\pm$ 2     & 115       & $\pm$ 26    & [m\,s$^{-1}$]         \\
$N_{\rm meas}$       & 22~(E)    & + 14~(S)    & 17~E      & + 4~S       &                       \\
$\Delta_{\rm S-E}$   & 23        & $\pm$ 4     & 93        & $\pm$ 17    & [m\,s$^{-1}$]         \\
$\sigma_{\rm E}$     & 17.5      &             & 12.1      &             & [m\,s$^{-1}$]         \\
$\sigma_{\rm S}$     & 9.0       &             & 4.1       &             & [m\,s$^{-1}$]         \\
$\sigma_{\rm ES}$    & 10.0      &             & 7.5       &             & [m\,s$^{-1}$]         \\
\noalign{\smallskip}
\hline
\noalign{\smallskip}
$a_{\rm 1} {\sin i}$ & 4.17      &             & 9.73      &             & [$10^{-4}$ AU]        \\
$f(m)$               & 7.06      &             & 1.30      &             & [$10^{-9}$ M$_\odot$] \\
$m_2 {\sin i}$       & 2.49      &             & 5.61      &             & [${\rm M_{Jup}}$]     \\
$a$                  & 0.24      &             & 1.98      &             & [AU]                  \\
\noalign{\smallskip}
\hline
\end{tabular}
\end{table}

\begin{figure}[t!]
\centering
  \begin{minipage}[t]{0.45\textwidth}
    \centering
    \resizebox{0.98\hsize}{!}{\includegraphics{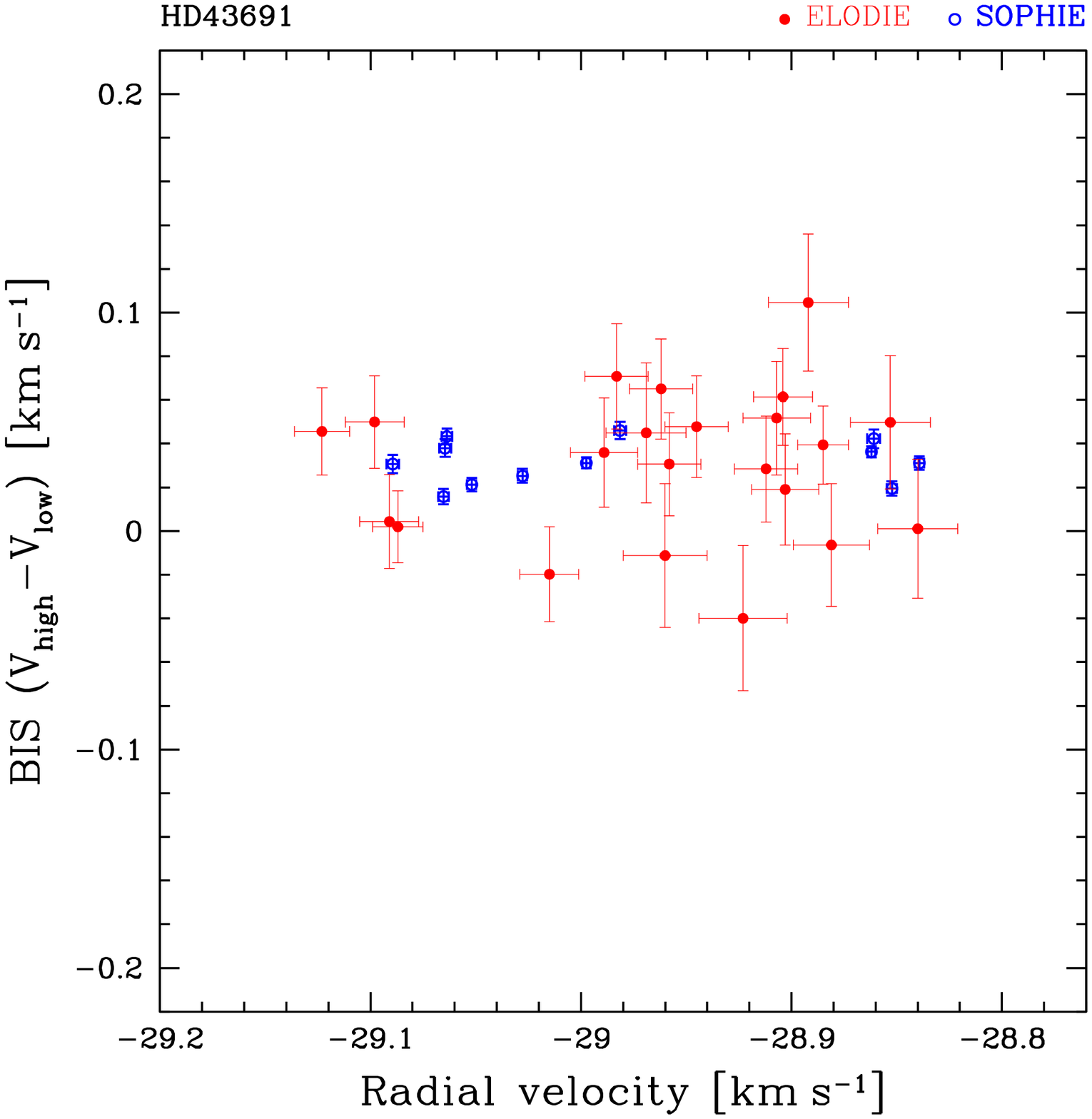}}
  \end{minipage} \\
  \begin{minipage}[t]{0.45\textwidth}
    \centering
    \resizebox{0.98\hsize}{!}{\includegraphics{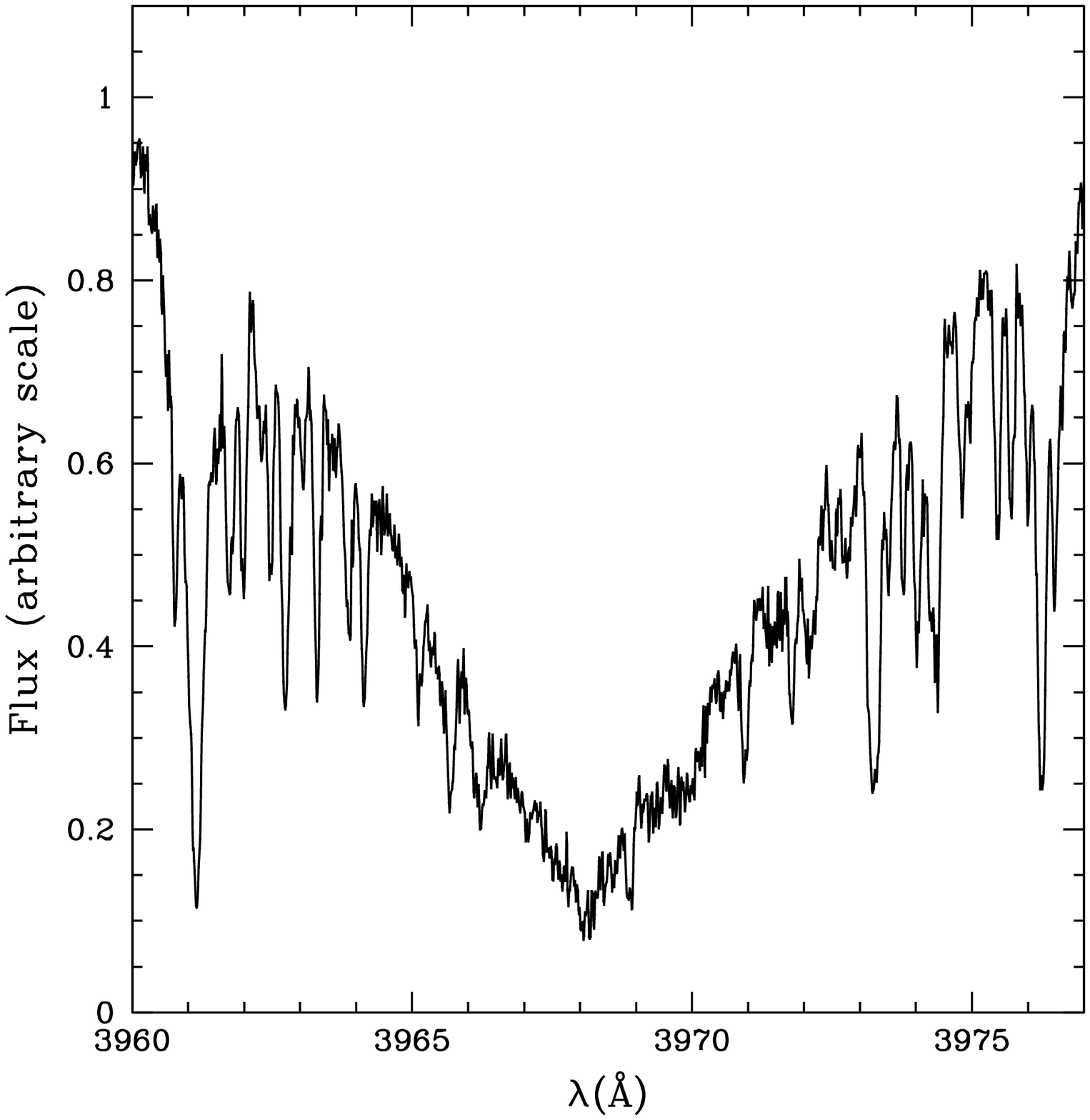}}
  \end{minipage}%
  \caption{\top Comparison between bisector inverse slope (BIS) and radial
           velocities of HD\,43691 showing no correlation between them
	   for both ELODIE and SOPHIE measurements.
	   \bot $\lambda$3968.5~\AA~\ion{Ca}{ii} absorption line region
	   of the high S/N spectrum obtained for HD\,43691. No clear
	   emission feature is observed in the center of this line,
	   indicating a low activity level.}
  \label{bis_ca}
\end{figure}

In the top panel of Fig.~\ref{rv_1} we plot the radial velocities of
HD\,43691 and the Keplerian fit adopted using the two sets of measurements.
The middle panel shows the residuals around the solution. The weighted rms
around the solution is $\sigma_{\rm E}$~= 17.5~m\,s$^{-1}$ for ELODIE, 
$\sigma_{\rm S}$~=~9.0~m\,s$^{-1}$ for SOPHIE, and
$\sigma_{\rm ES}$~=~10.0~m\,s$^{-1}$ for the whole dataset. The bottom panel
shows the phase-folded radial velocities. Table~\ref{orb_elem} lists the
adopted orbital elements, together with the inferred planetary parameters.

%
%
\subsection{Low chromospheric activity for HD\,43691}
The radial velocity variations observed for a star can also be the result of
physical events in the stellar atmosphere rather than the presence of an
orbital companion. For example, spots on the surface of an active star can
change the observed spectral-line profiles and induce periodic variations
of the measured radial velocities. By analysing the line-bisector
orientations one can distinguish which of these situations is the real
origin of the variations \citep{Quelozetal2001}.

The analysis of the line-bisector orientations, or bisector inverse slope
(BIS value), of HD\,43691 shows that there is no correlation between the BIS
values and the radial velocities derived (Fig.~\ref{bis_ca}, top panel).
Thus the observed variations in radial velocity are not induced by stellar
activity and rotation
\citep[as is the case of HD\,166435 in][]{Quelozetal2001}.
The observed behaviour of the line bisectors also indicates that the
radial velocity variations are not resulting from contamination by the light
of a late-type binary companion
\citep[see e.g. the case of HD\,41004 in][]{Santosetal2002}. Such variations
are thus most probably due to the presence of a planetary companion
orbiting HD\,43691.

The chromospheric activity level can also be verified by means of the
reemission in the core of \ion{Ca}{ii} absorption lines (e.g.
$\lambda$3968.5~\AA). By observing the respective spectral region in the
high S/N spectrum of HD\,43691 obtained with SOPHIE (Fig.~\ref{bis_ca},
bottom panel), we can note that this star is not active. Since this line is
located in the blue part of the spectral domain, where the flux is
appreciably lower, this kind of inspection requires a high S/N spectrum,
which is much better achieved by the higher efficiency of SOPHIE.

%
%
\section{A long-period planet orbiting HD\,132406}
\label{star_par2}
\subsection{Stellar characteristics of HD\,132406}
\label{star_car2}
HD\,132406 (HIP\,73146) is listed in the Hipparcos catalogue as a G0 star
with visual magnitude V = 8.45, color index $B-V$ = 0.65 and parallax $\pi$
= 14.09 $\pm$ 0.77 mas (71 pc distant from the Sun). These parameters set a
value of $M_V$ = 4.19 for the absolute magnitude. The bolometric correction
is BC = $-$0.062.

\begin{figure}[t!]
  \centering
  \resizebox{0.9\hsize}{!}{\includegraphics{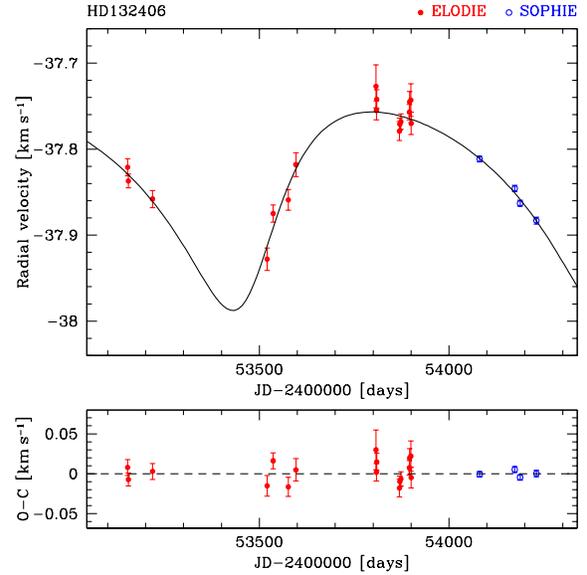}}
  \caption{\top ELODIE and SOPHIE radial velocities of HD\,132406 plotted
           together with the adopted Keplerian solution that better fits the
	   combined measurements.
	   \bot Residuals around the solution, with
	   $\sigma_{\rm E}$~=~12.1~m\,s$^{-1}$ for ELODIE,
	   $\sigma_{\rm S}$~=~4.1~m\,s$^{-1}$ for SOPHIE, and
	   $\sigma_{\rm ES}$~=~7.5~m\,s$^{-1}$ for the combined data points.
	   Error bars represent the photon-noise errors.}
  \label{rv_2}
\end{figure}

The metallicity and the projected rotation velocity of this star are
respectively [Fe/H] = 0.18 $\pm$ 0.05 and $v{\sin i}$ = 1.7 km\,s$^{-1}$,
estimated from the calibrations of the ELODIE cross-correlation functions.
The effective temperature derived is $T_{\rm eff}$ = 5885 $\pm$ 50~K and
comes from the calibrations of $T_{\rm eff}$ as a function of $B-V$ and
[Fe/H] from \citet{Santosetal2004}. The derived mass and age are $M_\star$ =
1.09 $\pm$ 0.05~M$_{\odot}$ and 6.4 $\pm$ 0.8~Gyr, from the Geneva models of
stellar evolution. These parameters are listed in Table~\ref{stellar_par}.
%
%
\subsection{Orbital solution for HD\,132406\,b}
A Keplerian solution fitted to the radial velocity measurements of
HD\,132406, from both ELODIE and SOPHIE observations, results in an orbit
with period $P$~=~974 $\pm$ 39~days and eccentricity $e$ = 0.34 $\pm$ 0.09.
The velocity zero-point of the two datasets is a free parameter, and the
difference between them is $\Delta_{\rm S-E}~=~93~\pm~17$~m\,s$^{-1}$.

The top panel of Fig.~\ref{rv_2} shows the radial velocities of this star
together with the best Keplerian solution. The bottom panel of the same
figure shows the residuals around the adopted solution, for which the
weighted rms
is $\sigma_{\rm E}$~=~12.1~m\,s$^{-1}$ for ELODIE,
$\sigma_{\rm S}$~=~4.1~m\,s$^{-1}$ for SOPHIE, and $\sigma_{\rm ES}$ =
7.5~m\,s$^{-1}$ for the combined set of measurements. HD\,132406 is slightly
fainter than HD\,43691, but has smaller photon-noise errors, which is
probably due to
broader line profiles of HD\,43691. Table~\ref{orb_elem} lists the orbital
elements and the planetary parameters of the HD\,132406 system, which was
obtained
with the best Keplerian fit.

\begin{figure}[t!]
  \centering
  \resizebox{0.9\hsize}{!}{\includegraphics{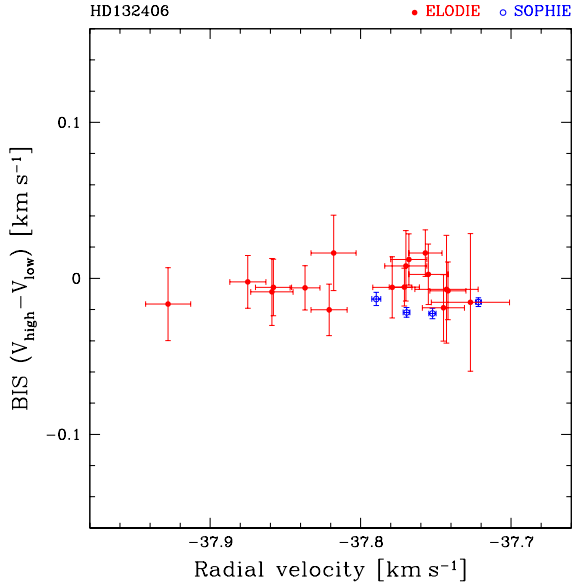}}
  \caption{Comparison between bisector inverse slope (BIS) and radial
           velocities of HD\,132406 showing no correlation between them.
	   Both ELODIE and SOPHIE measurements are plotted.}
  \label{bis}
\end{figure}

As in the case of the star HD\,43691, the analysis of the bisector inverse
slope of HD\,132406 shows no correlation between the BIS values and the
observed radial velocities (Fig.~\ref{bis}). In addition, no chromospheric
reemission is observed in the core of the \ion{Ca}{ii} absorption line at
$\lambda$3968.5~\AA.

%
%
\section{Discussion and Conclusions}
\label{conc}
In this paper we have announced the discovery of two new planet candidates
as results from our ELODIE search programme biased towards metal-rich stars.
In this programme, a total of six planets were discovered so far, out of
which four are hot Jupiters ($P < 10$~days) and two are the
intermediate-period planets just presented. Five of the host stars have
metallicity greater than 0.1~dex, while the one with a transiting very hot
Jupiter (HD\,189733) is a solar-metallicity star.

So far, we have observed at least once almost 82\% of the 1061 sample stars,
and for each one we estimated a value for metallicity. We thus verified that
about 26\% of the observed stars have [Fe/H] $\ge$ 0.1 dex (about 15\% for
$0.1\le$~[Fe/H]~$<0.2$, 8\% for $0.2\le$~[Fe/H]~$<0.3$ and 3\% for
$0.3\le$~[Fe/H]~$<0.4$). Furthermore, according to the percentage of stars
with planets per metallicity bin determined by \citet{Santosetal2004},
10$-$30\% of the stars with [Fe/H] $\ge$ 0.1~dex are likely to host a giant
planet (about 9, 24 and 28\% respectively for the same three ranges of
metallicity mentioned above). Applying these percentages to the 867 observed
stars, we finally find that the number of giant planets we expect to
discover in each of those metallicity ranges is respectively 12, 17 and 7, a
total of
36 planets, from which roughly 25\% (9 planets) are predicted to be hot
Jupiters.

Although most of our target stars were already observed, only 75\% of the
metal-rich stars have a minimum of three measurements. Stars with one or two
spectra need more observations before being rejected or classified as
possible planet hosts. On the other hand, long-period planets are more
difficult to detect, and stars showing long-term radial-velocity trends
also need more observations. In any case, we have already found almost a
half of the expected number of hot Jupiters among the metallic portion of
our sample.

The stars presented in this paper were also observed using the new SOPHIE
spectrograph. The proposed orbital solutions, firstly found with ELODIE,
were improved with the new observations. With ELODIE decommissioned, this
new instrument will also continue monitoring other high-metallicity stars,
especially the most promising cases.

\begin{acknowledgements}
  We thank the Swiss National Science Foundation (FNSRS) and the
  Geneva University for their continued support to our planet-search
  programmes, and the Observatoire de Haute Provence for the
  granted telescope time. N.C.S. would like to thank the support from
  Funda\c{c}\~ao para a Ci\^encia e a Tecnologia (FCT), Portugal, in the
  form of a grant (reference POCI/CTE-AST/56453/2004). This work was
  supported in part by the EC's FP6 and by FCT (with POCI2010 and FEDER
  funds), within the HELAS international collaboration. The support from
  Coordena\c c\~ao de Aperfei\c coamento de Pessoal de N\'\i vel Superior
  (CAPES - Brazil) to R.D.S. in the form of a scholarship are gratefully
  acknowledged as well.
\end{acknowledgements}
\bibliographystyle{aa}
\bibliography{Dasilvaetal2007}

\end{document}